\documentstyle[12pt]{article}
\begin{document}
\newcommand{\preprint}[1]{\begin{table}[t]  
           \begin{flushright}               
           \begin{large}{#1}\end{large}     
           \end{flushright}                 
           \end{table}}                     

\baselineskip 18pt

\preprint{TAUP-2087-93}
\title{Time Measurement in Quantum Gravity}
\author{N.Itzhaki\\Raymond and Beverly Sackler Faculty of Exact
Sciences\\School of Physics and Astronomy\\Tel Aviv University, Ramat Aviv,
69978, Israel}

\maketitle

\begin{abstract} We discuss time measurement in quantum gravity.
Using general relativity for large distances and the uncertainty principle we
find a minimum time interval of the order of the Planck time,
therefore the uncertainty in time measurement is bounded from below.
\end{abstract}

\newpage

In this letter we discuss a Gedanken experiment for the measurement of time and
show the existence of a minimal observable time interval in quantum gravity.
This result, in itself, is not completely new;
the existence of a minimal observable length has already been proposed in the
context of string theories in refs.\cite{ve,ko},
the fact that the concept of horizon is not defined at scales smaller than
Planck scale has been shown in \cite{ma}.
Our new points are : 1) we do not consider strings, but use general properties
of quatum gravity:
the uncertainty principle and Schwarzchild solution at large distances.
2) we find that the minimum error in time measurement is not a constant of
nature (Planck time) as might be expected, but a {\em function} of the distance
between the observer and the observed event.

In order to measure time, one must have a clock located at a distance $x$ from
the observer.
The observer obtains his information by looking at the clock, therefore the
clock must emit at least one photon toward him.

There are three causes of error in this process of time measurement:

1. The clock's accuracy---every clock has a minimum error $\Delta t$.

2. The time it takes the photon to reach the observer has uncertainty due to
the uncertainty of the metric caused by the clock's energy uncertainty $\Delta
E$.

3. The size of the clock---the uncertainty in the distance that the photon had
to travel in order to reach the observer is $2R$ ($R$ is the clock's radius),
therefore this error contributes $\frac{2R}{c}$ to the total error.

Classically there is no problem with this time measurement process,
since  we can eliminate those causes of error simultaneously,
then the time it takes the photon to reach the observer is \(\frac{x}{c}\)
 , and the clock's time is exactly the difference between the time the photon
reaches the observer and the time it takes the photon to reach him.
The uncertainty principle is about to change the whole picture as we shall see.

In the absence of a theory of quantum gravity we do not know the law of
gravitation at short distances but at large distances
(relative to the Planck length) general relativity would be a good
approximation to quantum gravity.

Suppose $R\le x_{c}$ ($x_{c}$ is the shortest distance for which we assume that
general relativity is a good approximation to quantum gravity,
thus $x_{c}=\alpha \sqrt{\frac{G\hbar}{c^{3}}}$) then the third cause of error
will countribute $F(R)$,
in the absence of a theory of quantum gravity all we can say about $F(R)$ is
that
$F(x_{c})=\frac{2}{c}x_{c}$ and $F(R)>0$ for $x_{c}>R>0$.

For distances larger than $x_{c}$ we can use general relativity, thus for
$r>x_{c}$ Schwarzschild solution would be a fair assumption.
For those distances we have
\begin{equation}ds^{2}=-c^{2}dt^{2}(1-\frac{2GE}{c^{4}r})+\frac{dr^{2}}{1-\frac{2GE}{c^{4}r}}\end{equation}
where $E$ is the energy of the clock, therefore the speed of light is
\begin{equation}v=\frac{dr}{dt}=c-\frac{2GE}{c^{3}r}\end{equation}
The time it takes the photon to reach the observer from $x_{c}$  is
\begin{equation}T=\int_{x_{c}}^{x}\frac{dr}{v}=\frac{1}{c}(x-x_{c})+
\frac{2}{c^{5}}GE\log\frac{c^{4}x-2GE}{c^{4}x_{c}-2GE}\end{equation}
Notice that $\frac{2}{c^{4}}GE<x_{c}$, otherwise the photon will be locked at
the clock's black hole.
We can use \begin{equation}\log\frac{x-a}{y-a}>\log\frac{x}{y},
(x>y>a>0)\end{equation}
to obtain \begin{equation}\Delta T> 2\frac{\Delta
E}{c^{5}}G\log\frac{x}{x_{c}}\end{equation}
Then we can use the uncertainty inequality \cite{ah} \(\Delta t\Delta
E\geq\hbar\) and $F(R)>0$
to obtain
 \begin{equation}\Delta T_{tot}(\Delta E)>\frac{\hbar}{\Delta E}+\frac{2\Delta
EG\log\frac{x}{x_{c}}}{c^{5}}\end{equation}
where \(\Delta T_{tot}\) is the error for the whole process.

Eq.(6) implies that there exists a minimum error
\begin{equation}\Delta
T_{min}=2\sqrt{\frac{2}{c^{5}}G\hbar}\sqrt{\log\frac{x}{x_{c}}}\end{equation}
at
\begin{equation}\Delta E=\sqrt{\frac{\hbar
c^{5}}{2G\log\frac{x}{x_{c}}}}\end{equation}
Notice that as mentioned above $\Delta E<c^{4}\frac{x_{c}}{2G}$ thus the
relation in eq.(7) is satisfied only for \(x>e^{\frac{2}{\alpha^{2}}}x_{c}\)

If $x_{c}<x<e^{\frac{2}{\alpha^{2}}}x_{c}$ then we obtain the minimum at
\begin{equation}\Delta
T_{min}=\frac{x_{c}}{c}(\frac{2}{\alpha^{2}}+\log\frac{x}{x_{c}})\end{equation}
at \begin{equation}\Delta E=\frac{x_{c}c^{4}}{2G}\end{equation}

Suppose $R>x_{c}$ then we can use general relativity inside the clock, thus the
third cause of error will countribute $\frac{2}{c}R$ .
The time it takes the photon to reach the observer from $R$  is
\begin{equation}T=\int_{x_{c}}^{x}\frac{dr}{v}=\frac{1}{c}(x-R)+
\frac{2}{c^{5}}GE\log\frac{c^{4}x-2GE}{c^{4}R-2GE}\end{equation}
thus \begin{equation}\Delta T>\frac{2}{c^{5}}\Delta
E\log\frac{x}{R}\end{equation}
then  \begin{equation}\Delta T_{tot}(\Delta E,R)=\Delta t+\Delta
T+2\frac{R}{c}> \frac{\hbar}{\Delta E}+\frac{2}{c^{5}}\Delta
EG\log\frac{x}{R}+2\frac{R}{c}\end{equation}
$R>2\frac{\Delta EG}{c^{3}}$ (otherwise the photon will  be locked at the
clock's black hole)
therefore the function \begin{equation}f(R)=\frac{\Delta
EG}{c^{3}}\log\frac{x}{R}+R\end{equation}
is an increasing function, thus in order to measure time as well as possible we
should use a clock with $R=x_{c}$ then

 \begin{equation}\Delta T_{tot}(\Delta E)>\frac{\hbar}{\Delta E}+\frac{2\Delta
EG\log\frac{x}{x_{c}}}{c^{5}}+\frac{2}{c}x_{c}>\frac{\hbar}{\Delta
E}+\frac{2\Delta EG\log\frac{x}{x_{c}}}{c^{5}}\end{equation}
and  we are left with the same uncertainty as before $(R<x_{c})$.

Note that in the discussion above we use the most simple time measurement
process, any particles that will be added must necessarily increase the
uncertainty of the metric without decreasing
$\Delta t$ thus the total error will get larger.
Therefore it seems that the uncertainty mentioned above is a basic property of
nature.

\centerline{\bf ACKNOWLEDGMENT }
I would like to thank professor Y.Aharonov and professor A.Casher for helpful
discussions.

\end{document}